# Designing optical glasses by machine learning coupled with a genetic algorithm

Daniel R. Cassar[1], Gisele G. dos Santos, Edgar D. Zanotto

*Department of Materials Engineering, Federal University of São Carlos, São Carlos, SP, Brazil*

**Abstract**

   Engineering new glass compositions have experienced a sturdy tendency to move forward from (educated) trial-and-error to data- and simulation-driven strategies. In this work, we developed a computer program that combines data-driven predictive models (in this case, neural networks) with a genetic algorithm to design glass compositions with desired combinations of properties. First, we induced predictive models for the glass transition temperature ($T_g$) using a dataset of 45,302 compositions with 39 different chemical elements, and for the refractive index ($n_d$) using a dataset of 41,225 compositions with 38 different chemical elements. Then, we searched for relevant glass compositions using a genetic algorithm informed by a design trend of glasses having high $n_d$ (1.7 or more) and low $T_g$ (500 °C or less). Two candidate compositions suggested by the combined algorithms were selected and produced in the laboratory. These compositions are significantly different from those in the datasets used to induce the predictive models, showing that the used method is indeed capable of exploration. Both glasses met the constraints of the work, which supports the proposed framework. Therefore, this new tool can be immediately used for accelerating the design of new glasses. These results are a stepping stone in the pathway of machine learning-guided design of novel glasses.

---

1   Corresponding author: contact@danielcassar.com.br.

# 1. Introduction

Glass science and technology are currently experiencing an "artificial intelligence renaissance." Even though many of the tools being used are not new, the interface between data and glass sciences has never seen so much interest [1–17]. Naturally, the number of reports on machine learning-based property prediction of glasses has surged in the past three years due to the availability of powerful computational tools and hardware, and the recent licensing of the SciGlass database under a permissive license (https://github.com/epam/SciGlass)—which has approximately 400,000 entries on composition-properties of glasses. Moreover, the high correlation between composition and properties for inorganic non-metallic glasses makes using data-driven tools significantly easier for these materials than for polycrystalline materials.

Most of the tools and machine learning models reported so far have been focused on predicting a chosen property given a glass composition [2,4,8,12,17–19]. For new glass development, however, it is paramount to solve the *inverse* design problem, finding possible compositions that are predicted to have a desired set of properties. This inverse design problem cannot be solved by traditional machine learning methods alone, to the best of our knowledge. Still, it can be tackled by combining machine learning and optimization algorithms. In this context, Nakamura and co-authors [16] recently used Bayesian optimization coupled with Gaussian process regression to search for oxide glasses with high refractive indices. Still in the subject of optical glasses, Tokuda and co-authors [20] applied the knowledge obtained from trained ML models to dope a glass composition to obtain a material with a high refractive index and low Abbe number.

This work aims to propose and test a framework to solve inverse design problems for glass development using genetic algorithms. While the proposed framework is general, here it will be tested for designing new optical glasses. Neural networks will induce the predictive models used here.

# 2. Design trends in optical glasses

The growth of the smartphone market and the demand for increasingly smaller and better-defined security and car cameras have attracted significant attention and fostered optical glass research. There is enormous interest in obtaining increasingly smaller, thinner, and more efficient light transmission lenses. The recent technological advances in 4K and 8K applications and Virtual/Mixed Reality lenses are also fueling the field of optical glasses. A relevant review article by Peter Hartmann and co-authors [21] listed some of the hottest trends in optical glass research, which were: *i*) high refractive indices (1.7 or more), *ii*) a high Abbe number (60 or more), *iii*) high refractive indices and a high Abbe number, and *iv*) high refractive indices and a low Abbe number.

Glasses with high refractive indices are desired because they reduce the degree of spherical aberration and enable lens design with reduced dimensions, e.g., targeting smartphones and small car cameras. Glasses with a high Abbe number (low dispersion) are used in optical systems that require a low degree of chromatic aberration. Glasses with high refractive indices and a high Abbe number could significantly impact optical glass technology, as it would make it possible to obtain smaller and thinner lenses with less color dispersion, again targeting the smartphone market. Glasses with high refractive indices and a low Abbe number are used for color correction in specific optical systems [21]. Figure 1 shows a 2D histogram of the Abbe diagram for oxide glasses in the SciGlass database. This figure clearly shows the current property envelope of the optical properties of oxide glasses.

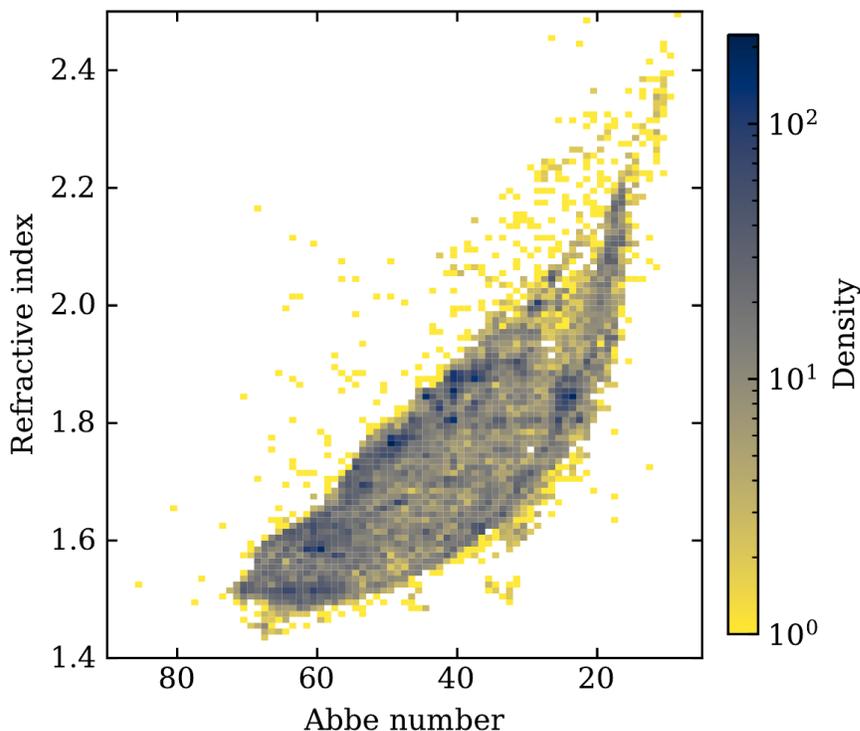

**Figure 1**. 2D histogram visualization of the Abbe diagram showing the refractive indices versus the Abbe number. Approximately 24,300 oxide glass data points were used to build this plot. Each rectangle has horizontal sides of 1, vertical sides of 0.01, and comprehends a density of experimental points depicted by its color (color bar on the right).

In addition to the four design trends previously discussed, there is additional interest in glasses having low glass transition temperature for optical lens production via precision molding techniques. This technique consists of applying pressure to a mold containing a glass-forming liquid, with a controlled atmosphere and temperatures between the glass transition temperature ($T_g$) and the softening point. The material obtained in this way is already in its final form, without the need for expensive, time-consuming additional steps, such as polishing and finishing. The most commonly used molds are made of tungsten carbide or silicon carbide with different coatings, and they are the most expensive parts of this process. Tungsten carbide molds, for example, are sensitive to oxidation at high temperatures, requiring the lenses to be conformed below 500 ºC, ideally below 450 ºC [21]. However, the same weak intermolecular bonds that allow a glass to have a low $T_g$ often bring some disadvantages, such as low chemical stability, which makes meeting this constraint a significant challenge.

# 3. Materials and methods

## 3.1. Data collection and partition

All the data used in this work were collected from the SciGlass database, which is now licensed under the ODC Open Database License (OdbL). This database collects glass properties and their respective chemical compositions reported in scientific articles, books, and patents.

We collected data on glass transition temperature and refractive index ($n_d$) of oxide glasses to induce predictive models for these properties via neural networks. The definition of oxide glasses considered here is the same as what we have used in previous work [10], that is: materials having an atomic fraction of oxygen of at least 0.3 and not having the chemical elements S, H, C, Pt, Au, F, Cl, N, Br, and I. Briefly, these are either elements that are too volatile or that occupy oxygen sites.

Both properties of interest depend on the thermal history and measurement conditions, which were not considered as features in this work as this information was not available in the database. While this lack of information increases the uncertainty of the prediction, it will be shown that induced models in this condition are still sufficiently accurate for the task at hand.

Before inducing the models, the dataset for each property was pre-processed following three steps: removal of glasses made with chemical elements having low representability, removal of glasses having properties with extremely low or high values, and replacement of duplicate entries by their median values. The *first* step is an iterative process where the fraction of examples containing each chemical element is computed, and then removing those glasses having chemical elements that are present in less than 1% of the examples. This process is repeated until all chemical elements are present in at least 1% of the examples. The rationale behind this choice is that each chemical element adds a new compositional dimension for the training of the model, and generalization may be compromised by having a small number of examples.

The *second* step is related to examples having extreme values of the glass properties, which were also removed. Here, extreme property values are defined as those below the 0.05% percentile and above the 99.95% percentile. The rationale behind this choice is the knowledge that the SciGlass database does not curate its entries, and a significant portion of typos or mistakes are located in these extreme regions.

The *third* and final step is related to examples with duplicate features, i.e., entries with the same nominal composition. These duplicate entries were grouped into a single entry with the median value of the property. The rationale behind this choice is to avoid a problem called data leakage [22], where the prediction of the model is artificially improved because it had "access" to data in the reserved dataset for testing. In other words, with duplicated data, it is possible that glasses having the same composition end up in different datasets; thus, information in the test dataset can "leak" into training.

We computed descriptive statistics on the datasets after collection and pre-processing. After this step, each dataset was partitioned into the *holdout dataset* (20%) and the *training and validation* (80%) dataset. The holdout set was not used for training the models nor for hyperparameter tuning; its main purpose was to measure the predictive power of the models. Finally, the final model used in the genetic algorithm optimization was trained using *all* the pre-processed data, as this is the usual practice for inducing the final predictive model.

## 3.2. Property prediction using neural networks

Neural networks (NN) are a group of machine learning (ML) algorithms that are excellent at finding patterns in data. They are the most used type of ML algorithm in the field of oxide glasses [1–4,6,8–13,15,17–19,23,24], and their success is due to their possibility of approximating any continuous function. The mathematical and statistical support for NNs is discussed in depth in the textbook by Charu Aggarwal [25].

This work investigates deep feedforward NNs with two hidden layers, which are reasonably good at predicting glass properties with acceptable precision [4,17]. One critical choice is the NN architecture because it is known that different problems often require different architectures. We used a hyperparameter tuning routine [26,27] to investigate some NN architectures, similar to previous work [4]. This process is described in detail in the Supplementary Material.

Each property of interest was investigated independently. In the end, we obtained a predictive model for each property. These models are functions for which the arguments lie in the chemical composition domain, and the output is a real number representing the predicted value of a given property. While the models can predict the properties of any glass with chemical elements within the domain of the functions, the expectation is that the prediction of compositions outside the training domain will lead to a much higher error.

## 3.3. Inverse design of glass compositions

Solving the inverse design problem, discussed in the introduction, requires an optimization algorithm. A genetic algorithm (GA) was chosen because it usually finds better solutions than random search and simulated annealing for problems with more than one optimal solution. Moreover, GA is trivially parallelizable, which may even yield better performance than simulated annealing in some cases.

GAs are meta-heuristics inspired by the theory of evolution and natural selection, bringing concepts of individuals, population, selection, reproduction, and mutation, and using them to navigate the multi-dimensional space of a single- or multi-objective optimization problem. More information on GAs is available in Koza's textbook [28] for a general view or Chakraborti's article [29] for a report focusing on materials design. To the best of our knowledge, the first published work to apply GA in the context of oxide liquids is that of Ojovan et al. [30], whereas the first to apply GA in the context of oxide glasses is that of Tandia et al. [8].

To use GA, one must first define how the "genome" of "individuals" are represented. In this work, an individual is defined as a glass with a particular chemical composition, having a genome that is represented as a row vector $\mathbf{I} = [x_1, x_2, \ldots, x_n]$, where each "gene" $x_i$ is an integer in the range [0, 100] that stores the amount (in moles) of a particular chemical compound $c_i$. Here, $n=28$ compounds were considered for the search space: $Al_2O_3$, $B_2O_3$, $BaO$, $Bi_2O_3$, $CaO$, $CdO$, $Gd_2O_3$, $GeO_2$, $K_2O$, $La_2O_3$, $Li_2O$,

MgO, Na$_2$O, Nb$_2$O$_5$, P$_2$O$_5$, PbO, Sb$_2$O$_3$, SiO$_2$, SnO$_2$, SrO, Ta$_2$O$_5$, TeO$_2$, TiO$_2$, WO$_3$, Y$_2$O$_3$, Yb$_2$O$_3$, ZnO, and ZrO$_2$. These compounds were selected because they are possible inputs for all predictive models trained here (see Table 3 in Section 4.1), i.e., they are common to both properties, and they were readily available in our laboratory for melting candidate compositions found by the GA.

A population **P** with $m$ individuals is defined as a matrix $m \times n$, where each row holds the information of a single individual. The population size of this work was $m=400$, and the initial population **P**$_1$ was generated by randomly sampling integers in the range of [0, 100] and building a 400×28 matrix. These randomly generated individuals are probably *inadequate* solutions to the optimization problem that is being investigated; many of them may not even form a glass or be within any constraint for which the problem is being optimized.

However, by pure chance, some randomly generated individuals of **P**$_1$ will be closer to a *possible* solution than others, even if they do not meet all the requirements of the problem. The word "possible" is emphasized because, for any given inverse design problem, a solution may or may not exist, which is an issue that is not directly related to GA.

The next step is to select the individuals of **P**$_1$ that will "survive" to the next generation and compose population **P**$_2$. This step starts by computing the fitness score for each individual. In this work, the fitness function $f$ was a weighted Euclidean distance in the property space, Eq. (1). The smaller the value of $f$, the better chances the individual has to survive.

$$f(x, y) = \sqrt{w_x(x - x_d)^2 + w_y(y - y_d)^2} + \varepsilon_1 + \varepsilon_2 + \varepsilon_3 \tag{1}$$

In the previous equation, $x$ and $y$ are the values of two different properties of a particular individual; $x_d$ and $y_d$ are the desired values for these two properties, which depend on the inverse design problem that is being solved; $w_x$ and $w_y$ are the weights that each property has to compute the fitness score; and $\varepsilon_1$, $\varepsilon_2$, and $\varepsilon_3$ are penalty factors that will be discussed later on in this section.

We studied only optimization problems with two properties, but Eq. (1) can be easily expanded for inverse design with more properties. To be clear, the values of $x$ and $y$ in Eq. (1) are predicted by the trained neural network models that were discussed in Section 3.2. Before the prediction, the chemical composition of each individual needs to be converted to atomic fraction and normalized to have a total sum of 1, as this was the format of the input data used for training the models. The weights $w$ used here were 1 for $T_g$ and 20 for $n_d$; these values were chosen to balance out the different magnitudes of these properties. This configuration was sufficient for this work, but poor convergence problems may benefit from tuning these weights.

After calculating the fitness score for all individuals in **P**$_1$, the selection phase begins. There are some selection strategies available. In this paper, we used the tournament selection where 3 individuals are selected at random from the population, and the one with the lowest fitness score from this group is selected to be part of the next generation, which is **P**$_2$ in this example. This process continues until **P**$_2$ has the same number of individuals as **P**$_1$.

The next phase is mating, where pairs of **P**$_2$ individuals can exchange genetic material, which replaces the original pair (the parents) with two new individuals (the offspring). The new individuals have a uniform chance of receiving each bit of genetic material from both parents, a process called uniform crossover. The chance of mating was set to 50%.

Finally, the last step of this iteration is the mutation phase, a critical step as it is the only opportunity for introducing different genetic material that was *not* present in the randomly generated $P_1$. Here, each individual of $P_2$ has a 20% chance to undergo mutation. If selected, then each gene has a 5% chance of changing its value to a random integer in the range of [0, 100]. On the one hand, if the mutation probabilities are too high, then the problem may not converge, as the "memory" of the best individuals is easily lost to mutations. On the other hand, if the mutation probabilities are too low, then the number of iterations required to reach a solution may become prohibitively large.

After these steps, the whole process is repeated by computing the fitness score and performing selection, mating, and mutation on $P_2$ to generate $P_3$. This iterative process was done until a solution was found or generation 5000 was reached.

We introduced two constraints for the GA search, one related to the minimum amount of glass-formers, and the other related to the chemical domain for which the predictive models were trained. Both constraints were computed independently for each individual. The first constraint checks for the ratio $\varphi$ between the sum of the glass network-forming oxides ($Al_2O_3$, $SiO_2$, $B_2O_3$, $GeO_2$, $P_2O_5$, $Sb_2O_3$, and $TeO_2$) and the total sum of compounds. If this ratio was below 45%, then a penalty $\varepsilon_1=(100(0.45-\varphi))^2$ was computed and considered in Eq. (1), otherwise $\varepsilon_1=0$. The rationale behind this constraint is to increase the chances that a composition found by the algorithm can be made into a glass. We know that this procedure does not guarantee that all compositions that meet this constraint can be vitrified by laboratory melt and quench techniques; however, it significantly increases the chances.

The second constraint checks if the composition is inside the chemical domain of the predictive models considered in the calculation of *f* in Eq. (1). For each chemical element *i* that is present in the individual, a distance $d_i$ is computed, which is zero if the atomic fraction of the said element is within the chemical domain of all the predictive models, or it is the absolute difference between the atomic fraction of the element and the closest atomic fraction within the domain of all predictive models. The penalty $\varepsilon_2=(100\sum_i d_i)^2$ is then computed for each individual. The rationale behind adding this constraint is that NNs trained using only the chemical composition as features are prone to higher prediction errors for compositions outside the domain. Sometimes, however, it may be desirable to explore chemical compositions close to the training domain, but not necessarily within it. Here, we relaxed the composition domain of each chemical element by 20%. It is important to stress that the chemical domain is only being checked by comparing the amount of each individual chemical element independently. This is different from another approach that is checking if the composition is inside the convex envelope considering *all* the chemical elements. A common misconception is that this penalty $\varepsilon_2$ would only allow the GA to find compositions that are far too similar to the training dataset of the predictive models, resulting in a procedure only capable of exploitation but not exploration. This is not true, as shown by the two glasses suggested by the algorithm (and produced here) being quite different from any composition in the available datasets (see the next section).

To have a clear difference in the fitness score between individuals that meet all constraints and individuals that do not, a final static penalty $\varepsilon_3$ is computed: if $\varepsilon_1 \neq 0$ or $\varepsilon_2 \neq 0$, then $\varepsilon_3=100$, otherwise

ε₃=0. The rationale behind adding this penalty is that we do not want to allow individuals outside the constraints of the problem (also known as infeasible individuals) to have even a small chance of winning the selection tournament against individuals that meet all the constraints. In other words, a death penalty when infeasible individuals compete against feasible individuals [31].

Finally, being a heuristic algorithm, GA is not guaranteed to reach a solution even if it exists. Any solution obtained is dependent on the randomly generated first population and the various steps that are due to chance. Because of this, the GA code was run many times to obtain a diverse set of solutions. The code used in this work was written in Python using the DEAP module [32], and it is available under the GPL3 license as the **GLAS** module [33], which stands for *Genetic Lookup for Apt Substances*.

## 3.4. Experimental tests

The design trend that guided our research was optical glasses with high refractive indices (1.7 or more) and low glass transition temperature (500 ºC or less). During an exploratory phase, we observed that some candidate glasses had poor chemical durability. Because of this problem, we manually reduced the search domain of the elements boron and phosphorous to [0, 0.02] and [0, 0.03] in atomic fraction, respectively. The rationale was that these two elements often decrease the chemical durability of glasses.

We obtained many composition candidates by running the GLAS software several times. We selected two glasses from the candidate list. The first was called Glass 1 and was the simplest glass from the list, that is, the one with the least amount of chemical compounds. The second was called Glass 2, and it was selected because it did not have an excessive amount of $ZrO_2$ and $Al_2O_3$ and a fair percentage of alkali and alkaline earth oxides to aid in the melting procedure. Their compositions are shown in Table 1, together with the target and predicted values of the properties of interest. We also checked how different these two selected glasses are from those present in the collected datasets (discussed in Section 3.1). To do so, we computed the Manhattan distance between the composition vector of Glass 1 and all the composition vectors of the glasses used for training the $T_g$ and $n_d$ models; the smallest this distance, the closer the chemical composition is. The glass shown in the "Closest to Glass 1" column in Table 1 is the one with the smallest distance. The same procedure was done for Glass 2; these analyses will be discussed later in the manuscript. The Manhattan distance was used instead of the Euclidean distance as the first performs better in high dimensional space [34].

**Table 1**. Composition (mol%), target, and predicted properties of the two glasses produced in this work. [†] To make this glass, we did not use MnO to avoid a strong color, instead we replaced it with ZnO. [‡] The uncertainty in the prediction is estimated by the RMSE value reported in Table 4.

| Oxide | Glass 1 | Glass 2 | Closest to Glass 1 | Closest to Glass 2 |
|---|---:|---:|---:|---:|
| $SiO_2$ | 66.67 | 41.75 | 81.08 | 50 |
| $Sb_2O_3$ | 21.21 | 27.18 | 18.92 | 40 |
| CaO | 3.03 | 1.94 | 0 | 0 |
| $B_2O_3$ | 3.03 | 0 | 0 | 0 |
| $Li_2O$ | 3.03 | 0 | 0 | 0 |
| $Nb_2O_5$ | 3.03 | 0 | 0 | 0 |
| $K_2O$ | 0 | 8.74 | 0 | 0 |
| $GeO_2$ | 0 | 7.77 | 0 | 0 |
| $Na_2O$ | 0 | 3.88 | 0 | 0 |
| $SnO_2$ | 0 | 2.91 | 0 | 0 |
| ZrO | 0 | 1.94 | 0 | 0 |
| MnO | 0 | 1.94[†] | 0 | 0 |
| ZnO | 0 | 0.97[†] | 0 | 0 |
| $La_2O_3$ | 0 | 0.97 | 0 | 0 |
| $Al_2O_3$ | 0 | 0 | 0 | 10 |
| **Target property** | **Glass 1** | **Glass 2** | | |
| Refractive index | 1.70 | 1.75 | | |
| Glass transition temperature (°C) | 450 | 400 | | |
| **Predicted property**[‡] | **Glass 1** | **Glass 2** | | |
| Refractive index | 1.71(3) | 1.76(3) | | |
| Glass transition temperature (°C) | 460(30) | 400(30) | | |

The reactants used and their respective purity are reported in the Supplementary Material. The chemicals were mixed, weighed, and homogenized in a rotation jar mill for 12 hours to make each glass. At the end of this process, the mixture was melted in a platinum crucible at a temperature range of 1000–1200 °C in a Deltech electric furnace, then poured over a metallic surface, crushed, and remelted for homogenization. This process was repeated three times. The melt was finally poured into a 1.5 × 1.5 × 3 cm graphite mold.

The glass transition temperature was determined for small pieces of the glasses by Differential Scanning Calorimetry (DSC, NETZSCH STA 449 F3 Jupiter), with a heating rate of 20 °C/min for Glass 1 and 10 °C/min for Glass 2. The refractive index was measured in 1.5 × 1.5 × 1.5 cm samples using the Na d-line (589.6 nm) of a Carl Zeiss Jena Pulfrich-refractometer PR2. Two adjacent faces of the samples (those that interacted with the light beam in the refractometer) were ground using 150–1200 mesh sandpaper and polished in velvet fabric with an aqueous cerium oxide suspension. No sign of chemical attack was observed. The refraction angle was measured and converted to refractive index using a conversion table provided by the equipment manufacturer.

Finally, both Glass 1 and Glass 2 were powdered and analyzed via X-ray diffraction in an Ultima IV, made by Rigaku. Glass pieces were milled until their particle size was reduced to 20 μm; they were analyzed in the step-scan mode in the interval between 10 and 70 degrees, with steps of 0.02° and 1 second per step. The Cu K$_\alpha$ radiation was used; no sign of crystallization was observed.

# 4. Results and discussion

## 4.1. Data analysis

Figure 2 shows the histogram for the two datasets used to induce the predictive neural network models. The distribution of the refractive index values has a single mode, with a clear skew to the right. The distribution of the glass transition temperature also has a single mode but is not visually skewed. Table 2 shows the descriptive statistics of both distributions.

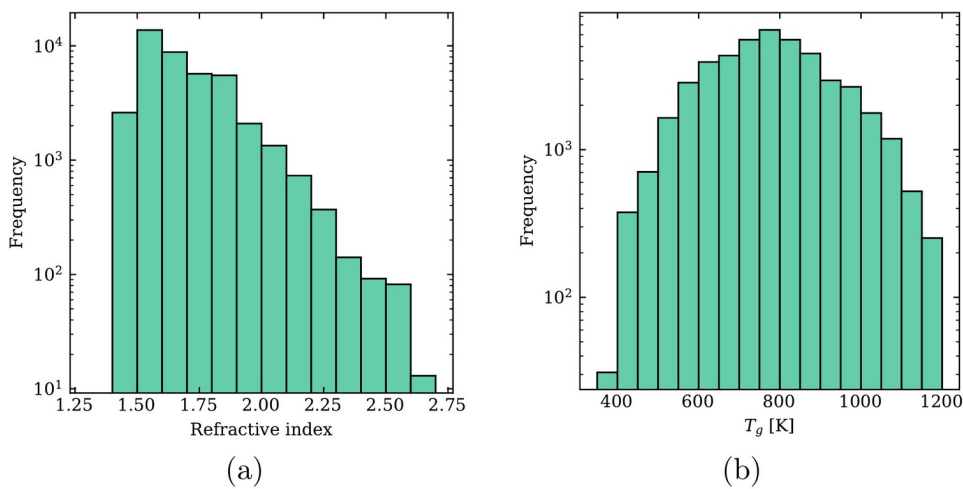

**Figure 2**: Distribution of the values of the (**a**) refractive index dataset and the (**b**) glass transition temperature dataset.

**Table 2**: Descriptive statistics of the datasets used for inducing the predictive neural network model. Glass transition temperature in Kelvin.

| Statistic | Refractive index | Glass transition temperature |
|---|---:|---:|
| Count | 41,225 | 45,302 |
| Number of chemical elements | 38 | 39 |
| Mean | 1.69 | 778.28 |
| Standard deviation | 0.18 | 150.56 |
| Minimum | 1.41 | 380.15 |
| Median | 1.64 | 773.15 |
| Maximum | 2.67 | 1271.15 |
| Skewness | 1.31 | 0.14 |
| Kurtosis | 2.11 | −0.33 |

Fig. 3 complements the analysis of the datasets by showing the distribution of examples with respect to the number of chemical elements. Glasses made with 4 chemical elements are the most common in both datasets, and multi-component glasses made with more than 10 elements are significantly less frequent than other multi-component glasses. Table 3 shows the chemical domain of both datasets, indicating the minimum and maximum atomic fraction for each element. As already mentioned, this information is relevant during the genetic algorithm search, as candidates that fall outside the *intersection* of chemical domains are penalized.

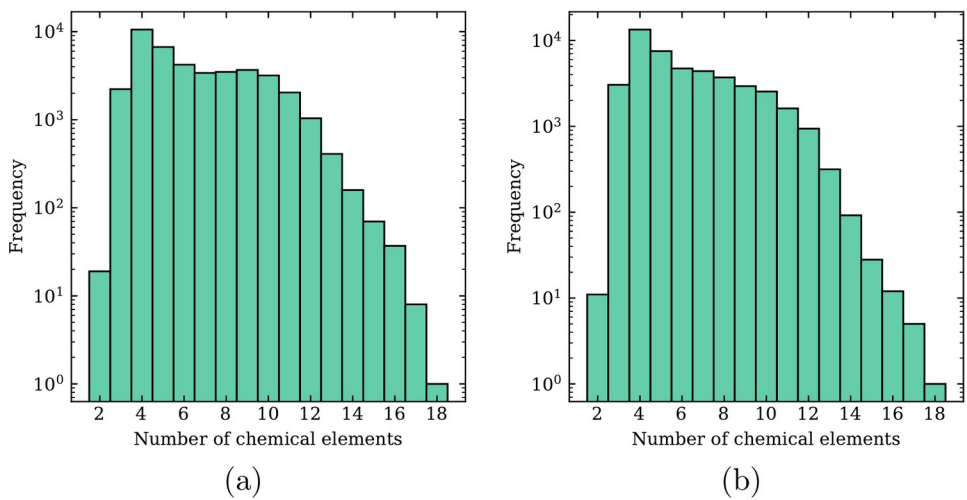

**Figure 3**: Distribution of the number of different chemical elements that make the glasses in the (**a**) refractive index dataset and (**b**) glass transition temperature dataset.

**Table 3**: Chemical domain for the refractive index and the glass transition temperature datasets in atomic fraction. Elements with a dash (–) are not present in the dataset.

| Element | Refractive index Min | Refractive index Max | Glass transition temperature Min | Glass transition temperature Max |
|---|---|---|---|---|
| Ag | – | – | 0 | 0.421 |
| Al | 0 | 0.38 | 0 | 0.367 |
| As | 0 | 0.4 | 0 | 0.4 |
| B | 0 | 0.4 | 0 | 0.4 |
| Ba | 0 | 0.326 | 0 | 0.244 |
| Be | 0 | 0.183 | – | – |
| Bi | 0 | 0.374 | 0 | 0.376 |
| Ca | 0 | 0.273 | 0 | 0.308 |
| Cd | 0 | 0.312 | – | – |
| Ce | – | – | 0 | 0.158 |
| Cs | 0 | 0.4 | 0 | 0.456 |
| Cu | – | – | 0 | 0.333 |
| Er | 0 | 0.171 | 0 | 0.143 |
| Fe | 0 | 0.222 | 0 | 0.316 |
| Ga | 0 | 0.4 | 0 | 0.334 |
| Gd | 0 | 0.4 | 0 | 0.179 |
| Ge | 0 | 0.376 | 0 | 0.385 |
| K | 0 | 0.418 | 0 | 0.497 |
| La | 0 | 0.4 | 0 | 0.255 |
| Li | 0 | 0.471 | 0 | 0.584 |
| Mg | 0 | 0.304 | 0 | 0.228 |
| Mn | 0 | 0.219 | 0 | 0.231 |
| Mo | – | – | 0 | 0.22 |
| Na | 0 | 0.553 | 0 | 0.553 |
| Nb | 0 | 0.26 | 0 | 0.263 |
| Nd | 0 | 0.175 | 0 | 0.2 |
| O | 0.379 | 0.739 | 0.316 | 0.745 |
| P | 0 | 0.286 | 0 | 0.286 |
| Pb | 0 | 0.437 | 0 | 0.442 |
| Sb | 0 | 0.4 | 0 | 0.4 |
| Si | 0 | 0.353 | 0 | 0.331 |
| Sn | 0 | 0.217 | 0 | 0.283 |
| Sr | 0 | 0.24 | 0 | 0.247 |
| Ta | 0 | 0.229 | 0 | 0.241 |
| Te | 0 | 0.333 | 0 | 0.338 |
| Th | 0 | 0.14 | – | – |
| Ti | 0 | 0.332 | 0 | 0.273 |
| V | – | – | 0 | 0.286 |
| W | 0 | 0.222 | 0 | 0.231 |
| Y | 0 | 0.4 | 0 | 0.188 |
| Yb | 0 | 0.4 | – | – |
| Zn | 0 | 0.286 | 0 | 0.321 |
| Zr | 0 | 0.232 | 0 | 0.202 |

## 4.2. Predictive models

Table 4 shows the hyperparameters used to induce the predictive neural networks. As expected, different problems often require different NN architectures, which is observed here by the architecture for predicting the refractive index that is reasonably different from the one to predict the glass transition temperature.

Table 4 also shows some metrics for the two models computed for the holdout dataset, which was not used for training the NNs nor during the hyperparameter tuning routine. Therefore, the metrics computed with this dataset simulate how the models behave with new unseen data. However, it is important to stress that the final predictive model used in the GLAS software was trained with *all* the available data, as our interest is to build the best predictive model within the considered framework. We expect that this final model will have a smaller prediction error than the one trained with only 80% of the dataset; however, by using the whole dataset to train the NN, we lost the ability to estimate the prediction errors. In other words, the expected errors of the final model are probably lower than the ones shown in Table 4, but we are unable to estimate them.

**Table 4**: Hyperparameters used to induce the predictive neural networks and metrics of the models. The hyperparameter tuning procedure is described in the Supplementary Material. Except for the $R^2$, all other metrics were rounded to one significant digit. [†] The numbers in parentheses refer to the hyperparameters of the first and second hidden layers, respectively. [‡] There are many ways to compute $R^2$; here, it was computed considering a linear model without an intercept as the alternative hypothesis.

| Hyperparameter | Refractive index | Glass transition temperature |
| --- | --- | --- |
| Activation function | ReLU | Sigmoid |
| Number of neurons[†] | (295, 115) | (190, 290) |
| Dropout[†] | (11%, 27%) | (8.2%, 25%) |
| Adam optimizer learning rate | $3.6 \times 10^{-4}$ | $1.3 \times 10^{-3}$ |
| Adam optimizer epsilon | $7.05 \times 10^{-7}$ | $2.57 \times 10^{-5}$ |
| Patience of the early stopping routine | 12 | 14 |
| Batch size | 256 | 128 |
| **Metrics (holdout dataset)** | | |
| Coefficient of determination[‡] ($R^2$) | 0.9997 | 0.998 |
| Relative deviation (RD) | 0.9% | 3% |
| Root mean squared error (RMSE) | 0.03 | 30 K |
| Mean absolute error (MAE) | 0.02 | 20 K |
| Median absolute error (MedAE) | 0.01 | 10 K |
| **Metrics (train & validation dataset)** | | |
| Coefficient of determination[‡] ($R^2$) | 0.9997 | 0.999 |
| Relative deviation (RD) | 0.8% | 2% |
| Root mean squared error (RMSE) | 0.03 | 30 K |
| Mean absolute error (MAE) | 0.01 | 20 K |
| Median absolute error (MedAE) | 0.009 | 10 K |

Neural networks can easily overfit the data due to the significant number of parameters in the model. This issue can happen even when hyperparameter tuning and cross-validation routines are employed, although the expectation is that these procedures help in reducing the problem. The best way to check for overfitting is by looking at the performance of the models on the holdout datasets; the performance reported here is comparable with state of the art in the field [10] and reasonable from an experimental point of view. Table 4 also shows the metrics for the training and validation datasets; they are only slightly better than the holdout dataset metrics, supporting that the final model is not overfitting the data. Figure 4 shows the loss for the train and validation datasets during the training of the NNs, showing no signs of overfitting (which would be a steep increase in the validation loss, meaning that the model is losing generalization power by fitting the noise in the data).

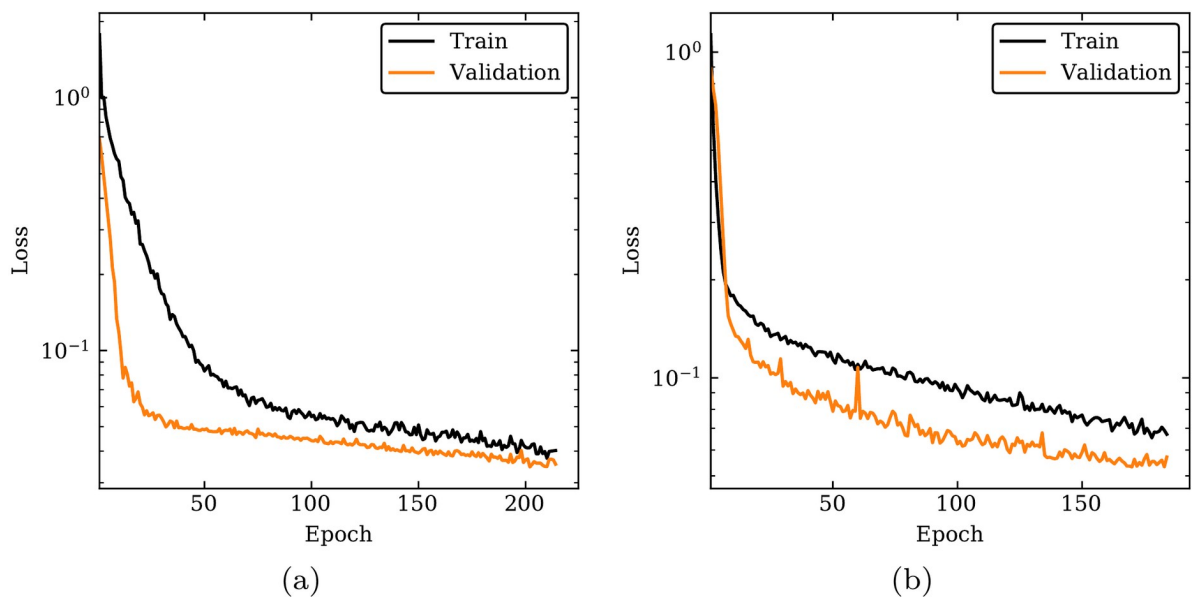

**Figure 4**: Loss curves for the training and validation datasets for (**a**) refractive index and (**b**) glass transition temperature. The validation loss is lower than the training loss because the dropout is not enabled during the validation phase. An epoch is when all the training data passes through the network during the training process.

Figure 5 complements the analysis of the predictive models by showing a correlation plot between predicted and reported values of the properties. These calculations were performed for the holdout dataset to understand how the predictive models behave in interpolating unseen data. Overall, most of the data points are close to the identity line, meaning that the predictions are reasonably close to the reported values. A more significant spread is observed in Fig. 5b when compared with Fig. 5a. One feature worth noting is the distribution of the prediction residuals for the refractive index, shown in the inset of Fig. 5a. This distribution is not symmetric, which merits further investigations.

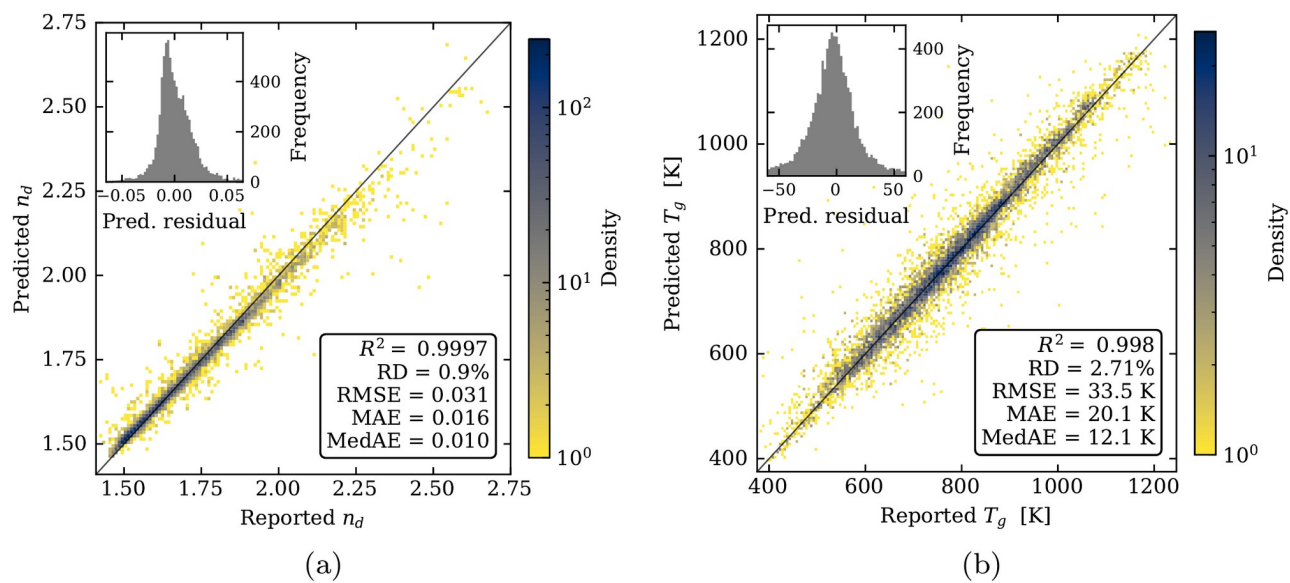

**Figure 5**: 2D histogram of the predicted versus reported values for (**a**) refractive index and (**b**) glass transition temperature, computed for the holdout dataset. The identity line is shown in black. The inset is the histogram of the prediction residuals, the difference between the reported and the predicted values. The vertical color bar shows the frequency of data points.

From a residual plot analysis (see Fig. S.1a in the Supplementary Material), we observe that the prediction residuals have a weak positive dependence with the predicted refractive index values, suggesting that additional features of glasses may improve the predictive power of the model. Chemical features can be easily extracted from the chemical composition, as described in detail by Ward et al. [35]. This procedure, however, will be left for future works as the current precision of the model (see Table 4) is sufficient for tackling the inverse design problem of this work.

The refractive index a reasonably linear property with respect to the chemical composition. In fact, a simple ordinary least squares linear regression (OLS) has a performance almost as good as the NN model, with an $R^2$ of 0.9993, RD of 1.3%, RMSE of 0.05, MAE of 0.02, and MedAE of 0.01. For investigations of other models to predict the refractive index see Refs. [17,36]. As expected, an OLS was not enough to capture the behavior of the glass transition temperature with respect of the composition, yielding a model significantly worse than the NN with $R^2$ of 0.994, RD of 6%, RMSE of 62 K, MAE of 45 K, and MedAE of 34 K.

The final analysis of the predictive models is the mean and standard deviation of the prediction residuals for each chemical element, shown in Fig. 6. Again, these calculations were performed for the holdout dataset. We observe from this figure that the prediction error depends on the chemical elements present in the glasses. Some chemical elements are susceptible to have data with more noise than others. Transition metals such as vanadium and volatile substances such as lead are examples of elements that can have additional noise in the data. Unaccounted changes in the charge of elements or the stoichiometry of the glass impact the non-crystalline structure, which governs many properties of glasses. The chemistry of some chemical elements such as erbium in Fig. 6a and arsenic in Fig. 6b was not captured by the respective models. In these cases, the data used for training the models was not representative enough for the model to generalize the element chemistry. All the chemical elements used to prepare Glass 1 and Glass 2 have an "average" standard deviation of the prediction residuals, except germanium when predicting the refractive index (only present in Glass 2), in which the standard deviation is high.

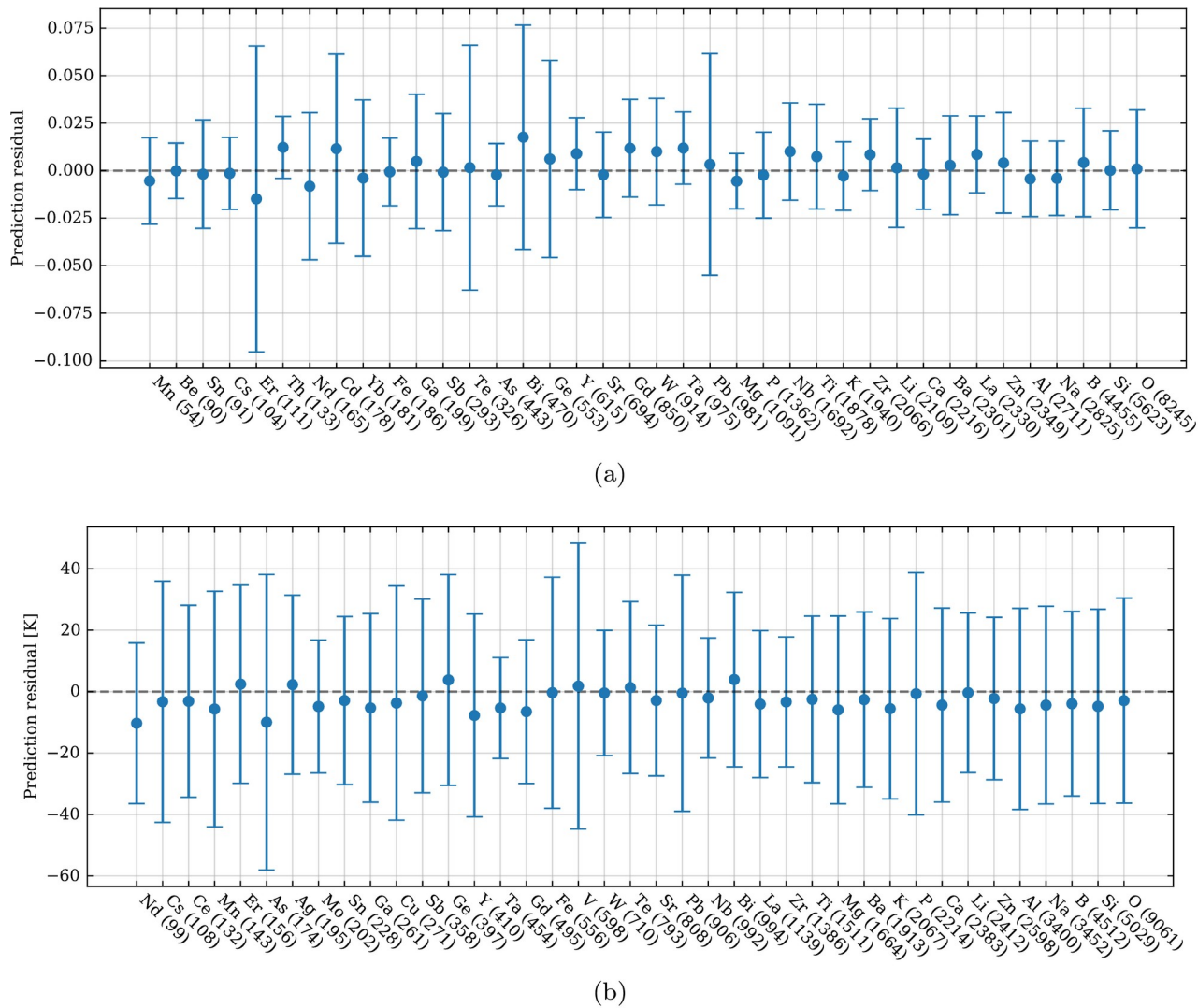

**Figure 6**: Mean and standard deviation of the prediction residual for each chemical element in the holdout dataset. (**a**) Refractive index and (**b**) glass transition temperature. The number in parentheses is the number of glass compositions containing the chemical element in the holdout dataset. The prediction residuals are the difference between the reported and the predicted values. The order of the elements is from the least to the most frequent, from left to right.

## 4.3. Experimental tests

The glass compositions we selected for the experimental tests (Table 1) contained a balanced amount of glass formers and other elements typically found in optical glasses. Both glasses made here are novel concerning the knowledge used to induce the predictive models—as can be seen by the closest composition in the training datasets being significantly different than the melted glasses (also shown in Table 1)—, showing that the proposed method is capable of exploration.

To make Glass 2, we replaced manganese oxide for zinc oxide due to the variation of oxidation numbers in the former that could give our glass strong colors. This procedure could've been easily implemented as a compound restriction in the GLAS search space as well. Even with this minor manual modification, our glasses end up showing a slightly yellowish color. This is likely because some elements, such as antimony and tin have variable valences. The viscosity of Glass 1 was considerably high, making it a challenge to obtain a homogeneous glass free of striae. This is not surprising when dealing with optical glasses; hence this glass required using a bar-shaped mold and a special casting technique to avoid cords.

The DSC traces used to measure the glass transition temperature are shown in Fig. 7, for which we obtained a value of 438(4) °C for Glass 1 and 450(13) °C for Glass 2. These analyses and computations of the uncertainty were done using software developed by Matthew Mancini [37]. The Glass 1 $T_g$ is within the predicted range of 460(30) °C; however, the Glass 2 $T_g$ is not within the predicted range of 400(30) °C, having a higher value than expected. $T_g$ is a tricky property to predict as besides the composition, it also depends on measurement conditions such as the heating rate of the experiment. Nevertheless, both glasses met the $T_g$ design trend that informed this work, with Glass 1 having the lowest value of $T_g$, an advantage over Glass 2.

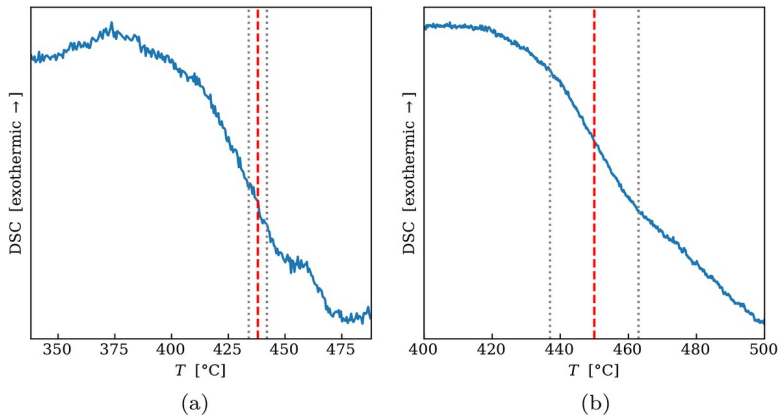

**Figure 7**. DSC traces focused on the glass transition region for (**a**) Glass 1 and (**b**) Glass 2. A linear baseline was subtracted for building the plots. The dashed red line shows $T_g$, and the dotted gray lines show the range of $T_g$ considering the uncertainty.

The measured refractive index for Glass 1 and Glass 2 were 1.713(1) and 1.749(1). Both are within the predicted value range of 1.71(3) for the first and 1.76(3) for the latter. It is important to mention that the predicted value for Glass 2 was for the *original* composition before the (minor) manual change that we discussed in the second paragraph of this section.

The obtained glasses met the material design that informed this work (refractive index above 1.7 and a glass transition temperature below 500 °C). All in all, these procedures, tools, and results are compelling, and we believe they are a stepping stone in the pathway of machine learning-guided design of new glasses for technological applications.

# 5. Summary and Conclusion

We developed a new computer program that couples data-driven predictive models with a genetic algorithm to aid in the design of new glass compositions with certain combinations of properties. As an example, after training predictive models for the glass transition temperature and refractive index, we searched for relevant glass compositions guided by a design trend regarding optical glasses—high refractive index and low glass transition temperature. Two candidate compositions suggested by the combined algorithms were selected and produced in the laboratory. The experimental properties of these glasses met the material design that informed this work, supporting the proposed framework.

This new tool can be immediately used for accelerating the design of new glasses, significantly minimizing trial-and-error. As it reduces the quantity of resources needed; it contributes to a greener glass development approach. These results pave the way for machine learning-guided design of novel glasses.

**Conflicts of interest**

There are no conflicts of interest to declare.

**Acknowledgments**

This study was financed by the São Paulo State Research Foundation support (FAPESP grant numbers 2017/12491-0 and 2013/07793-6), in part by the National Council for Scientific and Technological Development (CNPq, grant number: PQ 303886/2015-3 and 167434/2017-9), and in part by the Coordenação de Aperfeiçoamento de Pessoal de Nível Superior - Brasil (CAPES) - Finance Code 001. The Nippon Sheet Glass Foundation (NSG Foundation) Overseas grant is much appreciated. We would also like to thank our colleagues Ricardo Lancelotti and Dr. Laís Dantas, for helping with the refractive index and DSC measurements, respectively. We also thank Matthew Mancini for computing the glass transition values using his newly developed algorithm.

# Supplementary material

## 1. Chemicals used to produce the glasses

The chemical reactants used in this work are shown in the Table S.1. We used nitrates (when available) to create an oxidative atmosphere during the melting operation to control the oxidation numbers, as well as avoid chemical attacks on our platinum crucible.

**Table S.1**. Composition, manufacturer, and purity of the chemical reagents used in this work.

| Substance | Manufacturer | Purity |
|---|---|---|
| $SiO_2$ | Aldrich | >99.99% |
| $H_3BO_3$ | Vetec | 99.5% |
| $LiNO_3$ | Aldrich | 95% |
| $Ca(NO_3)_2.4H_2O$ | Vetec | 99% |
| $La_2O_3$ | Alfa Aesar | 99.99% |
| $Sb_2O_3$ | Aldrich | >99% |
| $GeO_2$ | Riedel-de-Haen | >99% |
| $KNO_3$ | Aldrich | >99% |
| $NaNO_3$ | Aldrich | >99% |
| $Nb_2O_5$ | CBMM | 98% |
| $SnO_2$ | Alfa Aesar | 99.90% |
| $ZnO$ | Riedel-de-Haen | >99% |
| $ZrO$ | Alfa Aesar | 99.70% |

## 2. Hyperparameter tuning

Hyperparameter tuning was done using the Python module hyperopt [26]. The search space of the hyperparameters are shown in Table S.2, and it was navigated with suggestions from a Tree-structured Parzen Estimator (TPE) algorithm [27]. A total of 150 hyperparameter sets were tested for each property of interest. The mathematical expressions of the activation functions that we explored are shown in Eqs. (S.1) to (S.3).

**Table S.2**. Search space of the hyperparameters of the neural networks. ReLU is the rectifier linear unit function and ELU is the exponential linear unit.

| Hyperparameter | Search space |
|---|---|
| Activation function | ReLU, ELU, or Sigmoid |
| Number of neurons in the first layer | [20, 300] |
| Number of neurons in the second layer | [20, 300] |
| Dropout probability of the first layer (%) | [0, 30] |
| Dropout probability of the second layer (%) | [0, 30] |
| Adam optimizer learning rate | [$10^{-4}$, $10^{-2}$] |
| Adam optimizer epsilon | [$10^{-7}$, $10^{-3}$] |
| Patience of the early stopping routine | [10, 14] |
| Batch size | 64, 128, or 256 |

$$\text{ReLU}(x) = \max{(0, x)} \tag{S.1}$$

$$\text{Sigmoid}(x) = \frac{1}{1+\exp(-x)} \tag{S.2}$$

$$\text{ELU}(x) = \begin{cases} x & \text{if } x \geq 0 \\ \exp{(x)} - 1 & \text{if } x < 0 \end{cases} \tag{S.3}$$

Before the hyperparameter tuning, the *training and validation* dataset was partitioned into 80% for local training, 10% for local validation, and 10% for local testing. Please note that the *training and validation* dataset was defined in Section 3.1 of the manuscript, and it *does not* contain the data in the *holdout* dataset. For each of the 150 sets of hyperparameters tested, a neural network was trained with this local training dataset and validated on the local validation dataset after each epoch. The validation step is important as the training stops if there is no improvement in the prediction of the validation dataset for a certain number of epochs defined by the patience hyperparameter. If this early stopping routine is never met, the neural network is then trained for 500 epochs.

Each of the 150 hyperparameter sets received a score value that is the mean squared error (MSE) of the prediction of the local test dataset. Those 10 sets with the lowest MSE score were tested again, this time in a 5-fold cross-validation analysis, where the hyperparameter set with the lowest average MSE score (considering all the folds) was the one selected to induce the final models. The selected sets of hyperparameters are shown in the main manuscript in the Table 4.

# 3. Residual plot analysis

Figure S.1 shows the residual plot analysis for the predictive models induced in this work.

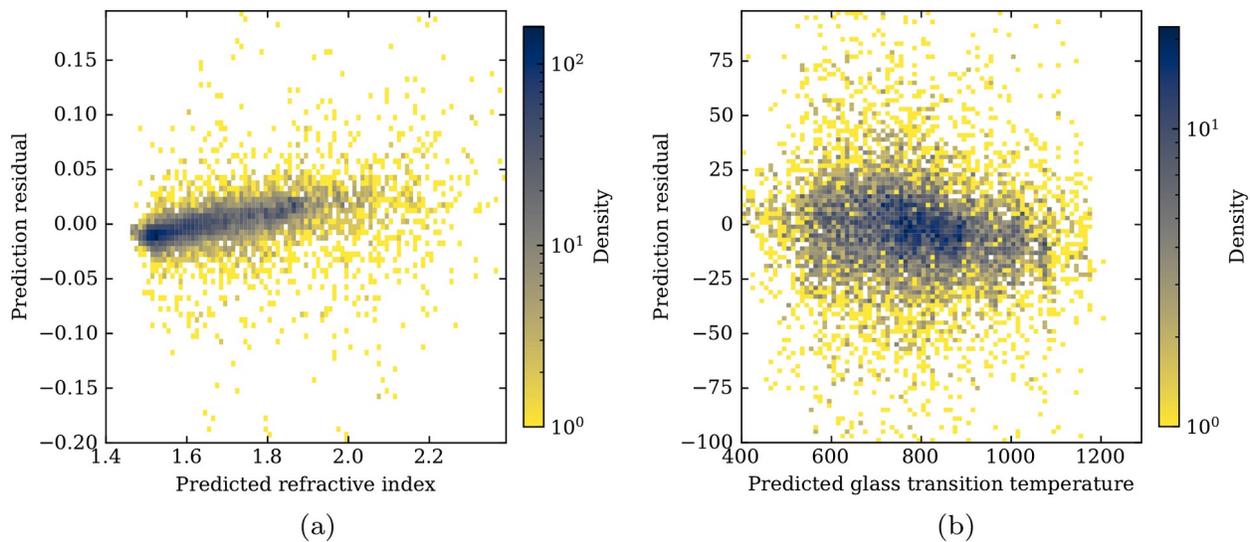

**Figure S.1**: 2D histogram of the prediction residual versus predicted (**a**) refractive index and (**b**) glass transition temperature, computed for the holdout dataset.